\documentclass[%
 amsmath,amssymb,
 reprint,%
]{revtex4-1}

\usepackage{graphicx}
\usepackage{dcolumn}
\usepackage{bm}

\usepackage[utf8]{inputenc}
\usepackage[T1]{fontenc}
\usepackage{mathptmx}
\usepackage{etoolbox}
\usepackage{xcolor}

\usepackage[colorinlistoftodos]{todonotes}

\makeatletter
\def\@email#1#2{%
 \endgroup
 \patchcmd{\titleblock@produce}
  {\frontmatter@RRAPformat}
  {\frontmatter@RRAPformat{\produce@RRAP{*#1\href{mailto:#2}{#2}}}\frontmatter@RRAPformat}
  {}{}
}%
\makeatother
\begin{document}

\preprint{AIP/123-QED}

\title[Research Article]{Desynchronous Learning in a Physics-Driven Learning Network}
\author{J. F. Wycoff}
\author{S. Dillavou}%
 \email{dillavou@upenn.edu}

\author{M. Stern}

\author{A. J. Liu}
\author{D. J. Durian}
\affiliation{ 
Department of Physics and Astronomy, University of Pennsylvania, Philadelphia, PA, USA 19104}

\date{\today}

\begin{abstract}

In a neuron network, synapses update individually using local information, allowing for entirely decentralized learning. In contrast, elements in an artificial neural network (ANN) are typically updated simultaneously using a central processor. Here we investigate the feasibility and effect of desynchronous learning in a recently introduced decentralized, physics-driven learning network. We show that desynchronizing the learning process does not degrade performance for a variety of tasks in an idealized simulation. In experiment, desynchronization actually \textit{improves} performance by allowing the system to better explore the discretized state space of solutions. We draw an analogy between desynchronization and mini-batching in stochastic gradient descent, and show that they have similar effects on the learning process. Desynchronizing the learning process establishes physics-driven learning networks as truly fully distributed learning machines, promoting better performance and scalability in deployment.
\end{abstract}

\maketitle

\section*{\label{sec:level1} INTRODUCTION}


Learning is a special case of memory~\cite{crowder_principles_2014,anderson_learning_2000}, where the goal is to encode targeted functional responses in a network~\cite{hopfield_neural_1982, mceliece_capacity_1987, rocks_designing_2017, stern_continual_2020}. Artificial Neural Networks (ANNs) are complex functions designed to achieve such targeted responses. These networks are trained by using gradient descent on a cost function, which evolves the system's parameters until a local minimum is found~\cite{lecun_deep_2015,mehta_highbias_2019}. Typically, this algorithm is modified such that subsections (batches) of data are used at each training step, effectively adding noise to the gradient calculation, known as Stochastic Gradient Descent (SGD) \cite{ruder_overview_2017}. This algorithm produces more generalizable results~\cite{keskar_improving_2017, chaudhari_stochastic_2018, smith_origin_2021}, i.e. better retention of the underlying features of the data set, by allowing the system to escape non-optimal fixed points~\cite{feng_inverse_2021, ruiz-garcia_tilting_2021}. This is reminiscent of noise-improving memory retention in physical systems such as sheared suspensions \cite{keim_generic_2011,paulsen_multiple_2014, keim_memory_2019}, where noise prevents the system from settling into equilibrium states where history-dependence is lost.  

Recent work~\cite{dillavou_demonstration_2021} has demonstrated the feasibility of entirely distributed, physics-driven learning in self-adjusting resistor networks. This system operates using Coupled Learning~\cite{stern_supervised_2021}, a theoretical framework for training physical systems using local rules~\cite{stern_shaping_2018, stern_supervised_2020, pashine_local_2021} and physical processes~\cite{pashine_directed_2019, hexner_effect_2020, hexner_periodic_2020} in lieu of gradient descent and a central processor. Because of its distributed nature, this system scales in speed and efficiency far better than ANNs and is robust to damage, and may one day be a useful platform for machine learning applications, or robust smart sensors. However, just like computational machine learning algorithms, this system (as well as other proposed distributed machine learning systems e.g.~\cite{scellier_equilibrium_2017, kendall_training_2020}) relies on a global synchronization of the learning rule, such that all elements change their resistance simultaneously. In contrast, the elements of the brain (neurons and synapses) evolve independently~\cite{abbott_synaptic_2000, kappel_network_2015}, suggesting that global synchronization is not required for effective learning. Desynchronizing the updates in machine learning is a largely unexplored topic, as doing so would be computationally inefficient. However in a distributed system such as the brain or self-adjusting resistor networks, it is the less restrictive modality~\cite{dolev_possibility_1986}, removing the need for a global communication across the network.

Here we demonstrate that desynchronous implementation of coupled learning is effective in self-adjusting resistor networks, in both simulation and experiment. Furthermore, we show that desynchronous learning can actually \textit{improve} performance by allowing the system to evolve indefinitely, escaping local minima. We draw a direct analogy between stochastic gradient descent and desynchronous learning, and show they have similar effects on the learning degrees of freedom in our system. Thus we are able to remove the final vestige of non-locality from our physics-driven learning network, moving it closer to biological implementations of learning. The ability to learn with entirely independent learning elements is expected to greatly improve the scalability of such physical learning systems.

\section*{Coupled Learning}
Coupled learning~\cite{stern_supervised_2021} is a theoretical framework similar to Equilibrium Propagation~\cite{scellier_equilibrium_2017,kendall_training_2020} that specifies evolution equations that enable supervised, contrastive learning in physical networks. In the case of a resistor network, inputs and outputs are applied and measured voltages at designated nodes of the network, and the edges self-modify their resistance according to local rules. The learning algorithm is as follows: Input and output nodes are selected, and a set of inputs from the training set is applied as voltages on the input nodes, creating the `free' response of the network. Using the measured outputs from this state $V^O_F$, the output nodes are then clamped at voltages $V^O_C$ given by
\begin{equation}
    \vec V_C^O = \eta \vec V^D + (1-\eta) \vec V_F^O
    \label{eta}
\end{equation}
where $V^D$ are the desired output voltages for this training example, and $0 < \eta \leq 1$ is an adjustable global parameter (``hyper parameter") that controls the strength of the nudge towards the clamped state. Thus, the output nodes are held at values closer to the desired outputs. When $\eta \ll 1$ this algorithm approaches gradient descent on a cost function~\cite{stern_supervised_2021}. This generates the `clamped' response of the network. The voltage drop across each edge in the free $\Delta V^F_i$ and clamped $\Delta V^C_i$ states then determine the coupled learning rule for changing the resistance of that edge:
\begin{equation}
    \delta R_i = \frac{\gamma}{R_i^2}\big([\Delta V^C_i]^2 - [\Delta V^F_i]^2\big)
    \label{originalcouple}
\end{equation}
where $R_i$ is the resistance of that edge and $\gamma$ is a hyper parameter that determines the learning rate of the system. In effect, this local learning rule lowers the power dissipation of the clamped state relative to the free state, nudging the entire system towards the (by definition) better clamped outputs. The system is then shown a new training example, and the process is repeated, iteratively improving the performance of the free state outputs. When a test set is given to the network to check its performance (by applying the input voltages appropriately) errors are calculated via the difference between the free state outputs and the desired outputs. A more detailed description of coupled learning is given in previous work~\cite{stern_supervised_2021}.

In the above algorithm, it is implicitly assumed that all edges update at the same time. Here we relax this assumption, modifying the learning rule Eq.~(\ref{originalcouple}) with a probabilistic element:
\begin{equation}
\Delta R_i(p) = \begin{cases} \delta R_i   \  & \text{with probability $p$}   \\   \ \  0  \ \  & \text{    otherwise} 
\end{cases}
\label{prob_update}
\end{equation}
where $0< p \leq 1$ is the update probability and $p=1$ recovers synchronized coupled learning. This modification, especially for low $p$, fundamentally changes how the system updates. Individual edges may spend long periods entirely static, while the system evolves around them, completely ignoring large changes along the way; that is, learning is desynchronized.

\begin{figure}
\includegraphics[width = \columnwidth]{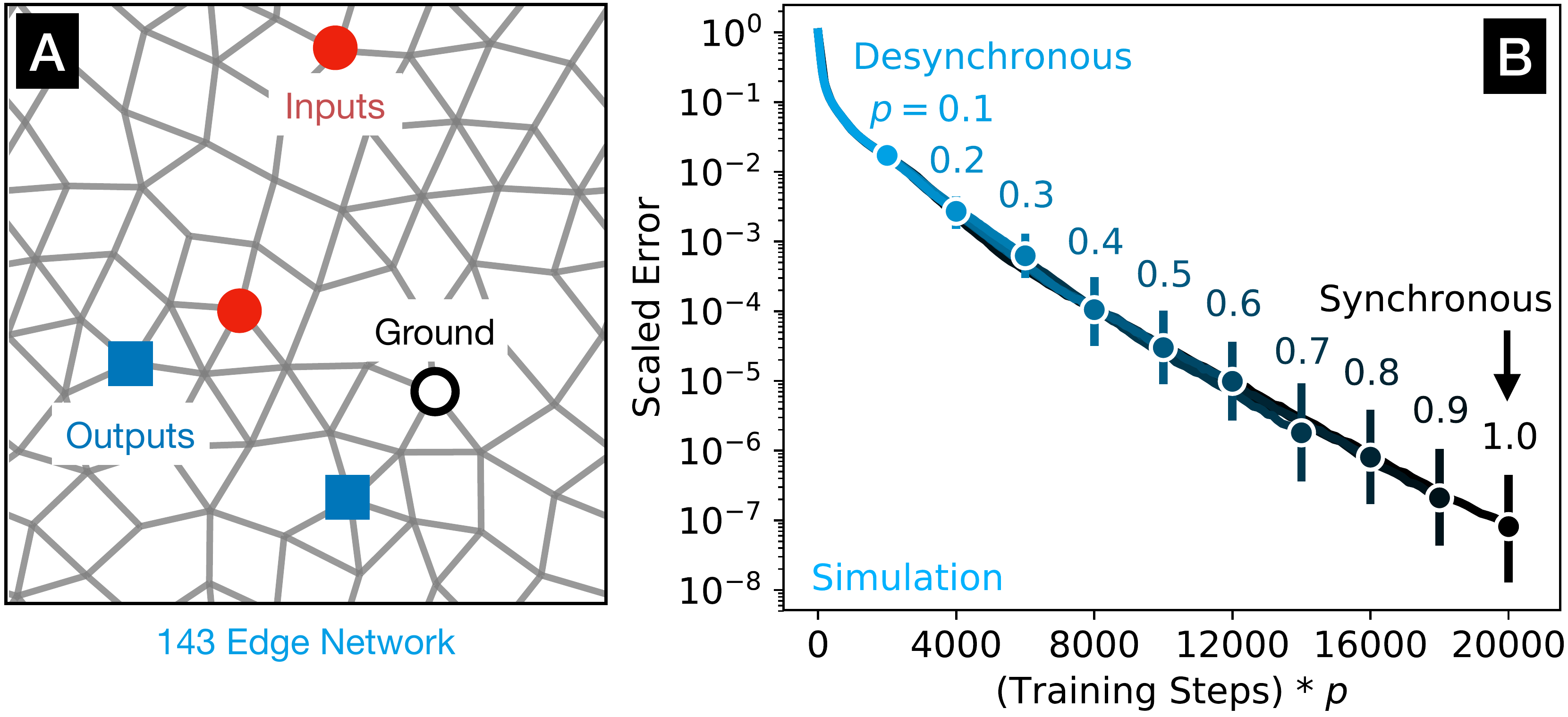}%
\caption{\textbf{Coupled Learning is Successful Without Synchronous Updates} \textbf{(A)} Simulated 143 edge coupled learning network. \textbf{(B)} Test set scaled error (error/error($t=0$)) curves averaged over 50 distinct 2-input 2-output regression tasks as a function of training steps times update probability $ p $. This x-axis scaling collapses the curves as each training step causes $143p$ edge updates, on average, proportionally changing the learning rate. Colors denote differing values of $ p $ ranging from $0.1$ to $1$. Error bars at the terminus of each curve denote range of final error values for a given $ p $ when run for 20000 steps. \label{fig1}}%
\end{figure}

Using simulations of coupled learning, per Ref.~\cite{stern_supervised_2021} but now with desynchronized updates, we find that the learning process is not hampered. In fact, the error as a function of training steps times $ p $ consistently collapses for all values of $p$ for a variety of tasks and networks, as shown for a typical example in Fig.~\ref{fig1}. This collapse occurs regardless of choice of hyper parameters $\eta$ (nudge amplitude) and $\gamma$ (learning rate). Notably, when updates become more desynchronous (decreasing $p$) solutions increasingly drift in resistance space from those found for synchronous learning (to be shown below in Fig.~\ref{fig4}A). These behaviors suggest that desynchronization may aid in exploring an under-constrained resistance space, much like stochastic gradient descent (SGD) in machine learning, a connection we now formalize mathematically.

\section*{Comparison to Stochastic Gradient Descent}

In computational machine learning, artificial neural networks can be trained using \emph{batch} gradient descent. In this algorithm, the entire set of training data is run through the network, and a global gradient is taken with respect to each weight in the network, averaged over the training set. The weights are then modified based on this gradient until a local minimum is found. In practice, this method is inefficient at best and intractable at worst~\cite{golmant_computational_2018}. A typical modification to this algorithm is known as stochastic gradient descent (SGD), where instead of the entire training set, a randomly selected subset of training examples (mini-batch) is used to calculate the gradient at each training step~\cite{ruder_overview_2017}. This effectively adds noise to the gradient calculation, speeds processing, and boosts overall performance by allowing the system to continually evolve, escaping from local minima in the global cost function. Stochastic gradient descent has been shown to improve learning performance in different settings, specifically in obtaining lower generalization (test) errors compared to full batch gradient descent. It is therefore argued that SGD performs implicit regularization during training, finding minima in the cost landscape that are more likely to generalize to unseen input examples~\cite{chaudhari_stochastic_2018}. 

This can be more clearly understood by describing training of a neural network as gradient descent dynamics of the learning degrees of freedom $\vec{w}$ (edge weights in a neural network) with an additional diffusion term, following Chaudhari \textit{et al.} \cite{chaudhari_stochastic_2018}. We define $\bar  b $ as the fraction of training data points used in a mini-batch. Full-batch ($ b  =1$) training simply minimizes the cost function $C(\vec{w})$, and thus the dynamics may be written as 
\begin{equation}
\gamma^{-1} d\vec{w}(t) = -\vec{\nabla}_w C(\vec{w}) dt 
\end{equation}
which yields solutions $\vec{w}_{\bar  b  = 1}$ that are minima of the cost function. When mini-batching, an additional diffusion term is added to the dynamics,
\begin{equation}
\begin{aligned}
\gamma^{-1} d\vec{w}(t) &= -\vec{\nabla}_w C(\vec{w}) dt + \sqrt{2 \gamma (\bar b  B)^{-1} D(\vec{w})} d\vec{W}(t)\\
D(\vec{w}) &= [B^{-1}\sum_i \vec{\nabla}_w C_i\otimes \vec{\nabla}_w C_i] - \vec{\nabla}_w C\otimes \vec{\nabla}_w C
\end{aligned}
  \label{eq:diffmatrix}
\end{equation}
where the diffusion matrix $D(\vec{w})$ is defined by outer products of the individual training example gradients, $B$ is the total number of training examples, and $dW$ is a Wiener process (random walk). These dynamics converge to critical points $\vec{w}_{\bar b }$ that are different from the minima of the cost function, $\vec{w}_{\bar b =1}$, by a factor that scales with the fraction of data points not included in each batch $(1 - \bar b )$. This difference is the hallmark of regularization, in this case performed implicitly by SGD.

\begin{figure}
\includegraphics[width = \columnwidth]{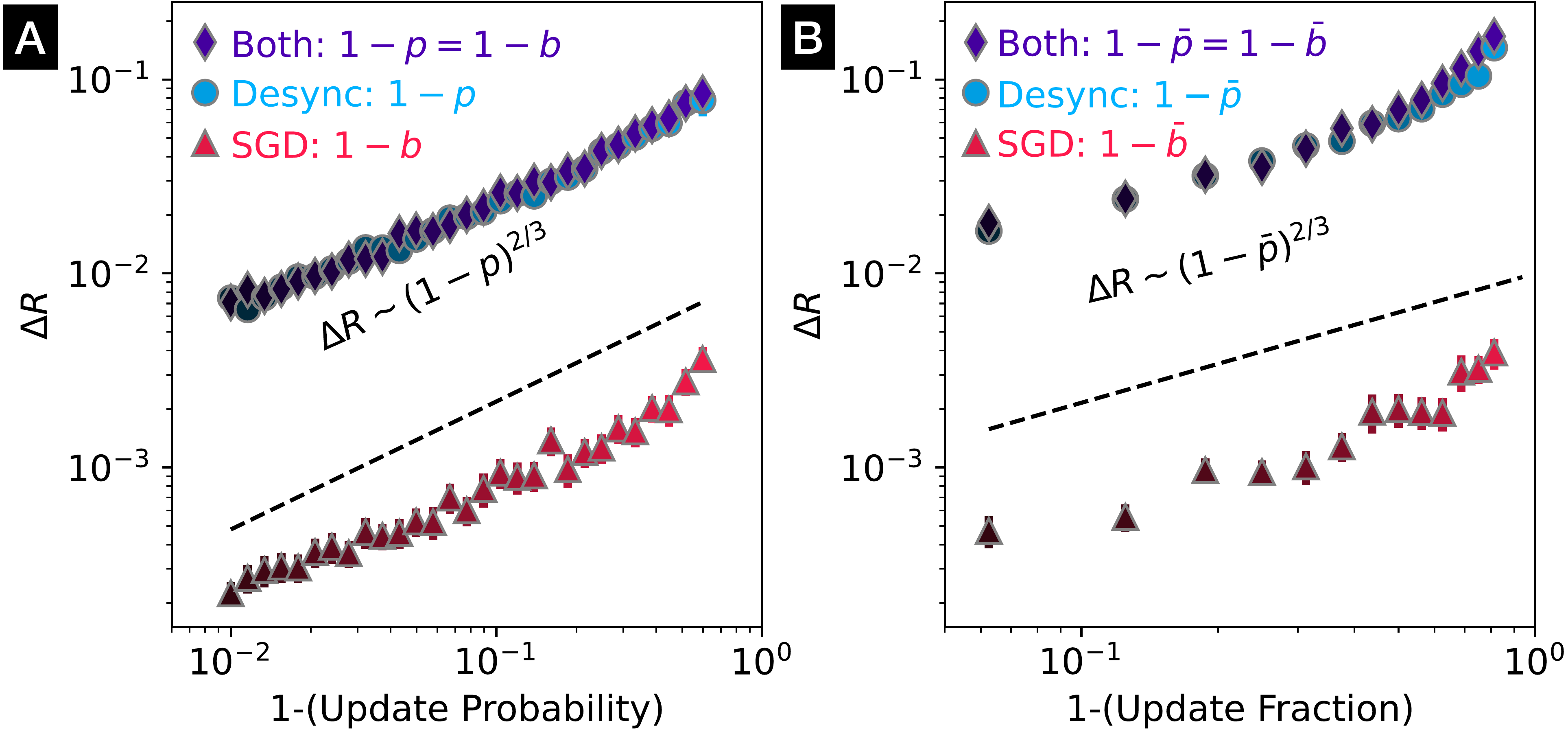}%
\caption{\textbf{Desynchronous Learning Behaves Like Stochastic Gradient Descent } \textbf{(A)} Distance in continuous resistor space from synchronized, full-batched solution as a function of $1- p $ for a 16-edge simulated, continuous network. Each data point represents an average over 50 regression tasks, each with 2 inputs and 2 outputs. Note mini-batching (stochastic gradient descent) and desynchronization generate the same power law, as does their combined effect. The vertical shift results from an effective learning rate difference. \textbf{(B)} Same as (A) but with constant number of edges updating or batch size (or both) at each training step.
\label{fig4}}%
\end{figure}

In coupled learning, the desynchronization of edge updates is expected to yield a similar effect. Instead of having different training examples, learning stochastically uses the gradient at independent edges. Therefore we can define an effective diffusion matrix for desynchronous coupled learning by
\begin{equation}
\begin{aligned}
\gamma^2 D_{eff}(R) &= [N^{-1}\sum_i \Delta R_i\otimes \Delta R_i] - \Delta \vec{R}\otimes \Delta \vec{R}
\end{aligned}
  \label{eq:diffmatrix2}
\end{equation}
where $N$ is the total number of edges. Note the similar form to the second line of Eq. (\ref{eq:diffmatrix}). With this definition, the analogy of desynchronous coupled learning and SGD is clear, with the edge update probability $ p $ playing the role of the batch fraction $\bar  b $, and thus we expect similar results for the two methods. We verify the analogy between desynchronous coupled learning and SGD in simulation.

For simulations with continuously variable resistors, we observe no change in final error when learning is desynchronized. This is consistent with expectations from SGD when tasks have large, multi-dimensional zero-error basins that are always found by the system. However, the analogy between SGD and desynchronization can still be explored by observing the solutions in resistor space. As a base case, we simulate a $N=16$ edge network (the same structure we will use in our experimental setup) using the original coupled learning rule (Eq.~\ref{originalcouple}) with a \textit{full batch} to solve a regression task with $B=16$ training examples. That is for a given edge $i$,
\begin{equation}
    \Delta R_i =\sum_{j=1}^{B}  \Delta R_{ij} = \sum_{j=1}^{B} \frac{\gamma}{R_i^2}\big([\Delta V^C_{ij}]^2 - [\Delta V^F_{ij}]^2\big)
    \label{fullbatch}
\end{equation}
where $j$ is the index of the training example, summed over all $B=16$ elements of the training set. This is an entirely deterministic algorithm, given initial conditions of $R_i$, and thus a good basis for comparison. Then we compare two forms of stochasticity, randomly choosing edges (desynchronization) and randomly choosing training examples (SGD). With probability $ p $ we update edges ($i$), and with probability $ b $ we include each training example in the sum ($j$). For $ b =1$ we use a full batch, and for $ p  = 1$ we update every edge synchronously. Coupled learning as described in previous work \cite{stern_supervised_2021,dillavou_demonstration_2021} used $ p =1$ and $ b  \ll 1$ (a single training data point at a time). Decreasing $ p $ (desynchronizing) and decreasing $ b $ (stochastic mini-batching) do not meaningfully change the final error of the network's solutions in continuous coupled learning, but do find \textit{different} solutions than the full-batch synchronous case. In fact, we find they have the same relationships to the fully deterministic solutions,
\begin{equation}
     b  = 1 : \ \ L_2\left(\vec R( p =1),\vec R( p )\right) \sim (1- p )^{2/3}
\end{equation}\begin{equation}
     p  = 1: \ \ L_2\left(\vec R( b =1),\vec R( b ) \right) \sim (1- b )^{2/3}
\end{equation}
Enforcing $ p  =  b $ also gives the same power law, all seen in Fig.~\ref{fig4}(A). We may also enforce a randomly selected but consistent fraction of edges ($\bar  p $) or of the training set ($\bar  b $) to be updated/included for each training step. This is the standard means of mini-batching in SGD, as mentioned previously. We find similar parallels between desynchronous and mini-batched learning in this condition, as seen in Fig.~\ref{fig4}(B). The overall multiplicative factor separating the data can be explained by SGD and the desynchronous learning rule having a different effective learning rate. Matching these effective rates collapses all data in Fig.~\ref{fig4}(A) and (B). 

This robust analogy between desynchronization and SGD suggests that in a system with a more disconnected cost landscape, we should expect error \textit{improvements} when desynchronizing coupled learning. We now turn to such a system, our experimental realization of a 16-edge network, where the resistor values are discretized, which decreases the number of degrees of freedom and prevents the system from settling into a minimum of exactly zero. As we will show, the experimental system successfully learns in the desynchronized regime, in some cases improving upon the synchronized solutions. Desynchronization thus allows a substantial simplification for implementation, especially in large networks, by removing the requirement for simultaneous updates across the entire system.

\begin{figure}
\includegraphics[width = \columnwidth]{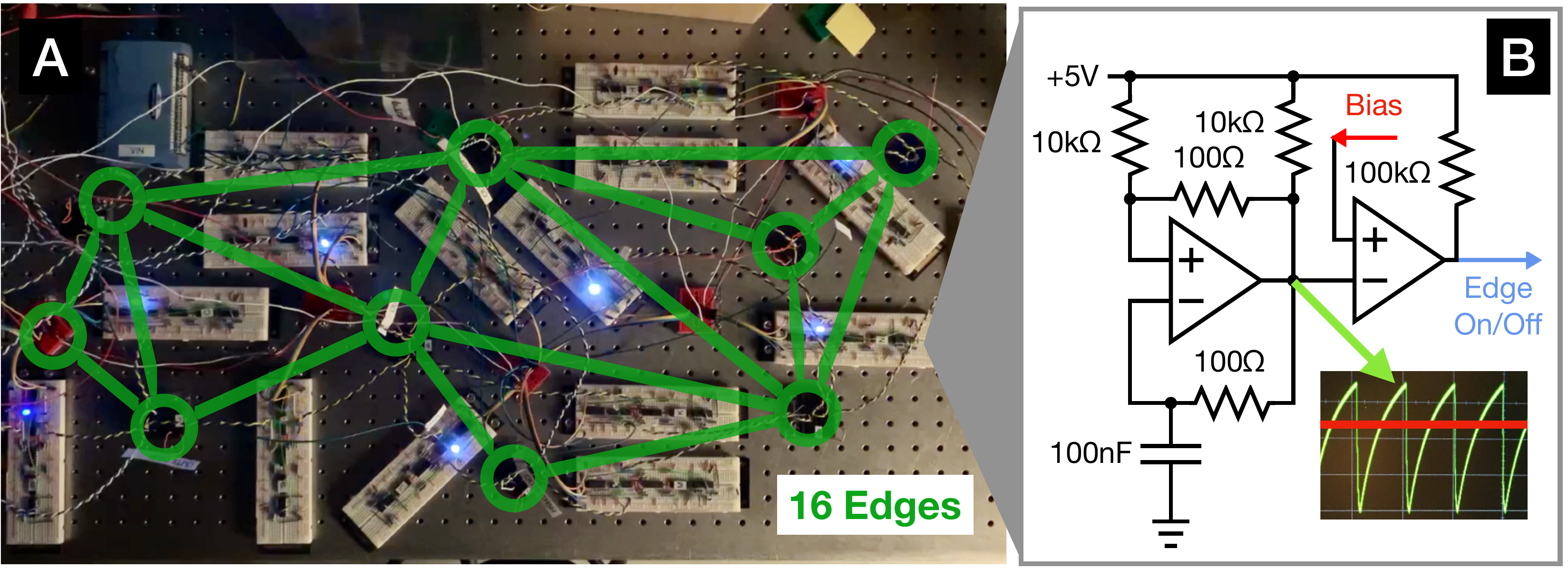}%
\caption{\textbf{Circuitry for Realization of Desynchronous Coupled Learning.} \textbf{(A)} Image of the entire 16-edge network. Edges with LEDs on are active (updating) on this training step. \textbf{(B)} Diagram of the oscillator circuit in each edge in (A). A global bias voltage (red) determines $ p $. Each edge compares the bias against against a local oscillator signal (green) to determine if its resistance is updated.
\label{fig2}}%
\end{figure}

\section*{Experimental (Discrete) Coupled Learning}
We test desynchronous updates in an experimental realization of coupled learning. In recent work \cite{dillavou_demonstration_2021}, coupled learning was first implemented in a physical system. In this system, contrastive learning was performed in real time by using two identical twin networks to access the free and clamped states of the network simultaneously. The system was robust to real-world noise, and successfully trained itself to perform a variety of tasks using a simplified version of the update rule that allowed only discrete values of $R$, specifically
\begin{equation}
    \delta R_i = 
     \begin{cases}
      +r_0 & \text{if} \ |\Delta V^C_i| +\sigma> |\Delta V^F_i| \\
      -r_0 & \text{otherwise.}
    \end{cases}  
    \label{signS}
\end{equation}
Note that we have explicitly added the measured bias of the comparators $\sigma$, which we find manifests as a random, uniformly distributed variable from 0 to 0.05~V. Previously, each edge in the network performed this update individually, but did so all at once, synchronized by a global clock. Here, we implement this learning rule~\footnote{Specifically in this work we use comparators and an XOR gate to evaluate XOR[ $(\Delta V^C_i > \Delta V^F_i)$ , $(\Delta V^C_i+\Delta V^C_i > 0)$].} but incorporate a probabilistic element, such that with probability $ p $ each edge updates according to Eq.~(\ref{signS}) on a given training step. Thus, we are able to tune the system from entirely synchronous $(p=1)$ to entirely desynchronous $ (p\ll1)$. 

We implement this  probabilistic functionality via separate circuits housed  locally with each twin edge of the network, shown in Fig.~\ref{fig2}(A).  This circuit, when triggered by a global signal, compares its local oscillating voltage signal to a global `bias' voltage, as shown in Fig.~\ref{fig2}(B). The components (comparators, capacitors, and resistors) used in each implementation of the oscillator vary slightly, changing the period  and phase of oscillation; thus the signals on each edge rapidly desynchronize.  In experiment, we find a Pearson correlation between pairs of edges to be consistently of order 0.01 for an update probability of 50\%, indicating that edges are updating independently. By changing the bias value, we can select  a wide range of values of $p$ for our experimental system.

\begin{figure}
\includegraphics[width = \columnwidth]{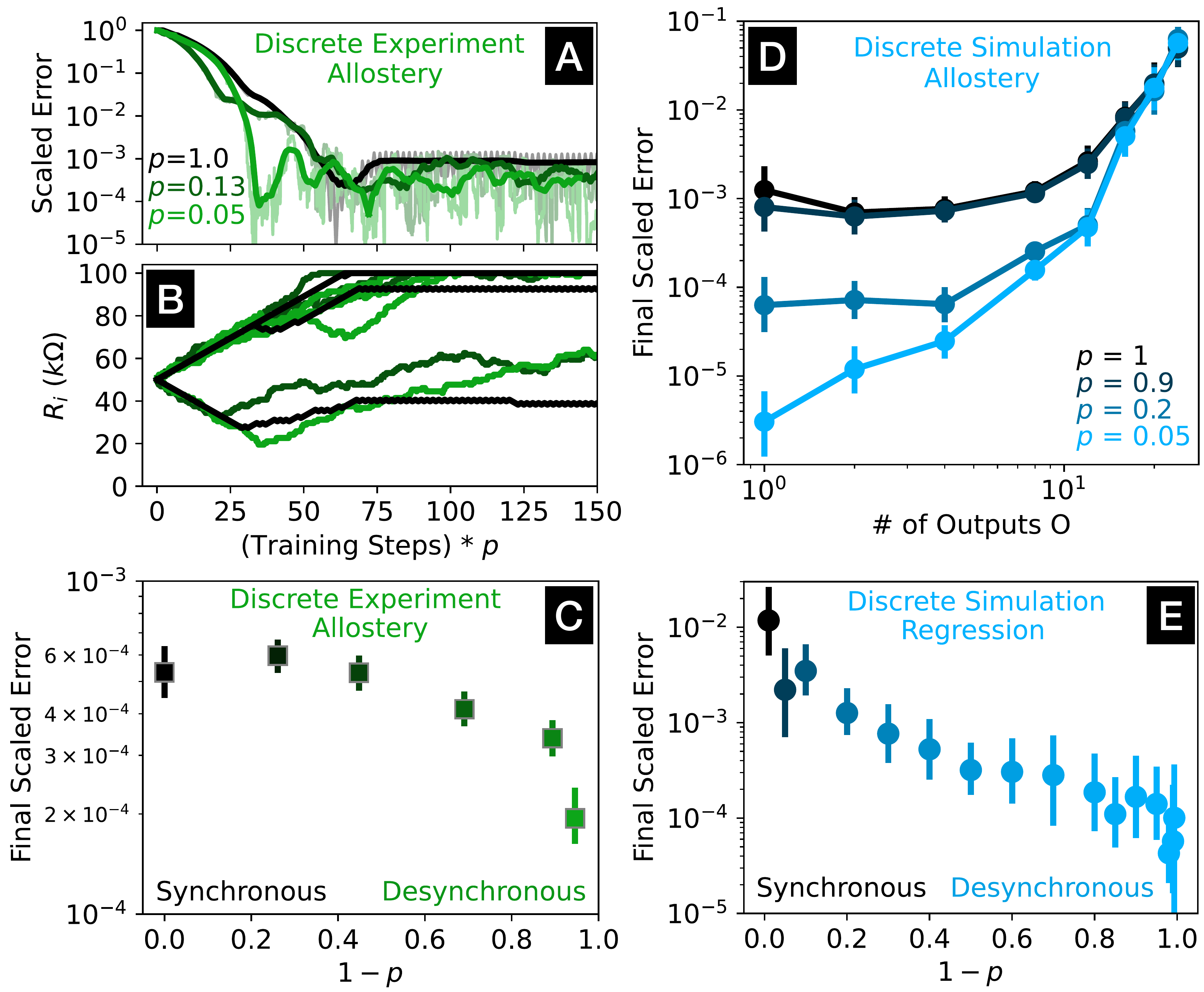}%
\caption{\textbf{Desynchronization Improves Discrete Network Solutions in Experiment and Simulation.} 
\textbf{(A)} Scaled error (error/error($t=0$)) vs training steps scaled by update probability $ p $ in experiment for an allosteric task  with 2 inputs and 2 outputs. One typical raw (faded) and smoothed (color) curve is shown for each of the three values of $ p $. 
\textbf{(B)} Three resistor values vs training steps scaled by update probability from the experiments shown in (A). 
\textbf{(C)} Scaled error at the end of training averaged over 25 allosteric tasks  each with 2 inputs and 2 outputs as a function of $ p $.
\textbf{(D)} Scaled error at the end of training for allosteric tasks as a function of number of outputs $O$. Each data point is an average over 20 tasks, each with $O$ outputs, $O$ inputs, and $O/2$ ground nodes, increasingly constraining the network as $O$ grows. Note the collapse of curves of varying $p$ as the task complexity grows. 
\textbf{(E)} Scaled test set error at the end of training in simulation averaged over 10 regression tasks  with 2 inputs and 2 outputs. In (D) and (E) the same 143 edge simulated network from Fig.~\ref{fig1}(A) is used with the discrete update rule (Eq.~\ref{simulation_discrete}).
 \label{fig3}}%
\end{figure}

As with the continuous version of coupled learning, desynchronization does not prohibit the discrete, experimental system from learning. In fact, desynchronized learning performs \textit{better} on average than synchronous learning for ``allosteric" (fixed input and output) tasks, as apparent in typical error curves as shown in Fig.~\ref{fig3}(A). Why does this stochasticity improve final errors? In short, it is because randomness allows the network to explore resistance space. Edges continually evolve when $ p <1$ (desynchronous), whereas for $ p =1$ (synchronous), the system may find a local minimum and remain there indefinitely, as shown by the flat black resistor traces in Fig.~\ref{fig3}(B). The ability to escape minima improves as the network becomes more desynchronized, leading to improved final error as $ p $ decreases for allosteric tasks in experiment, as shown in Fig.~\ref{fig3}(C). As tasks become too difficult, the beneficial effects of desynchronization are diminished. For a two-output, two-input regression task, our 16-edge experimental network shows no benefit from desynchronization. However, as we now show in simulation, increasing the size of the network brings learning back into a regime where desynchronization confers an advantage.

To test the advantages of desynchronous learning for future larger realizations, we perform a simulation tailored to match our experimental system but with more edges. We use the discrete update rule, Eq.~(\ref{signS}), limit our resistance values to 128 linearly spaced values and use $\sigma = U[0,0.05]$~V (uniformly sampled between 0 and 0.05~V).  As before, to desynchronize learning we have edges follow the update rule only with probability $p$ on each training step,
\begin{equation}
\Delta R_i(p) = \begin{cases} \delta R_i   \  & \text{with probability $p$}   \\   \ \  0  \ \  & \text{    otherwise} 
\end{cases}
    \label{simulation_discrete}
\end{equation}
That is, Eq.~(\ref{signS}) performed on each edge with probability $p$.
The addition of $\sigma$ leads to a tendency for the resistor values to drift upwards, just like in the experiment, finding lower power solutions, and putting the resistors in a regime where they can take smaller steps relative to their magnitude. From simulations of a 143-edge discrete network, we find that as allostery task complexity (number of both inputs and outputs, O,) increases, the beneficial effects of desynchronous learning diminish, as shown in Fig.~\ref{fig3}(D). More complex tasks require more desynchronous (lower $p$) learning to confer an advantage over synchronous learning. For tasks with enough outputs, moderately desynchronous learning yields indistinguishable error from synchronous learning, as shown by the overlap of the blue and black curves on the right of Fig.~\ref{fig3}(D). 

Unlike the experimental 16-edge network, desynchronization does improve the error for our simulated 143-edge learning a two-input two-output regression task, as shown in Fig.~\ref{fig3}(E). We believe that for such a task, our 16-edge experimental network is in the `too-complex' regime, whereas our simulated 143-edge network is not, and therefore shows a monotonic trend in final error with $p$.

Linear tasks like allostery and linear regression do not have local minima when the parameters in the linear kernel are free to change continuously~\cite{rencher_linear_2008}. In our networks the case is different, as the input-output relationship is always a linear function, but the linear kernel depends non-linearly on each resistance value, which are themselves the degrees of freedom. As a result, the cost landscape can have local minima. Even so, we see no evidence for local (non-zero) minima in our continuous simulations, likely because we have a very large number of degrees of freedom relative to the number of constraints.  In the discrete case, however, resistor space has fewer degrees of freedom, leading to more local minima that can trap the synchronous solution and preventing it from finding a global optimum.  Thus, desynchronizing the edges ultimately helps find deeper minima in the discrete system (Fig.~\ref{fig3}), but not in the continuous system (Fig.~\ref{fig1}) where we find no evidence of non-zero minima. 

\section*{Discussion}

In this work we have demonstrated the feasibility of learning without globally synchronized updates in a physics-based learning network, both with a continuous state space of solutions and a discrete one, in simulation and experiment. In all cases desynchronizing the learning process does not hamper the ability of the system to learn, and in the discrete resistor space with many local minima, actually improves learning outcomes. We have shown that this improvement likely comes from a behavior analogous to stochastic gradient descent, namely that injecting noise into the learning process allows the system to escape local minima and find better overall solutions. We have mathematically formalized this analogy and showed that mini-batching and desynchronization produce the same scaling of distance in solution space compared to a fully deterministic (full batch, synchronous) algorithm.

The freedom to avoid global synchronization is an important step towards total decentralization of the learning process in a physical system; it is necessary to make a \textit{learning material}. In this and previous \cite{dillavou_demonstration_2021} work, the experimental system is still run via a global clock, and thus requires a one bit communication with every edge to trigger resistor updates. However, the success at all values of $ p $ demonstrates that edges with entirely self-triggered updates should also function well. For a larger, less precise, tighter packed, or three-dimensional learning systems, removing this connection to each edge may greatly simplify construction. Furthermore, allowing desynchronization opens the door for learning with new types of systems that cannot be synchronized, such as elements updating out of equilibrium~\cite{stern_physical_2021}, or that include thermal noise~\cite{kappel_network_2015} or other stochastic processes.

In discrete-valued coupled learning, mini-batching alone (the standard in Coupled Learning) gives inferior results to mini-batching plus desynchronous updates. This suggests that in other learning problems with many local minima, including in artificial neural networks, desynchronous updates could benefit the learning process. While we are not aware of this desynchronization algorithm used in such a way, similar methods such as dropout~\footnote{In dropout, some fraction of edges in a layer of a neural network are removed for that training step. This is distinct from desynchronous learning, where all edges are present for calculating the outputs, but some simply do not update.} have been shown to be beneficial in improving generalizability of solutions~\cite{srivastava_dropout_2014}, similar to stochastic gradient descent. True desynchronization would be extremely inefficient in such a system, as then the entire gradient calculation is necessary for a single edge update. However, we have shown that benefits can be accrued by only moderate desynchronization, \textit{e.g.} 80\% update probability, which slows the learning process proportionately. The true test of the usefulness of this algorithm will be in larger, nonlinear networks solving problems on complex cost landscapes. This is a subject for future work.

\acknowledgements{Thanks to Marc Z. Miskin for insightful discussions, including on circuit design. This work was supported by the National Science Foundation via the UPenn MRSEC/DMR-1720530 (S.D. and D.J.D.) and DMR-2005749 (M.S.) and the Simons Foundation via Investigator Award 327939 (A.J.L.).}
\bibliography{bib}

\end{document}